\documentclass{jaa}

\usepackage{graphicx}
\usepackage{natbib}
\usepackage{ulem}
\usepackage{hyperref}
\usepackage{subcaption}
\hypersetup{colorlinks,linkcolor={black},citecolor={blue},urlcolor={red}} 

\begin{document}\sloppy

\title{Long-term spectroscopic monitoring of comet 46P/Wirtanen}

\author{K.Aravind\textsuperscript{1,*}, Kumar Venkataramani\textsuperscript{2,3}, Shashikiran Ganesh\textsuperscript{1}, Emmanuel Jehin\textsuperscript{4}, Youssef Moulane\textsuperscript{5}}
\affilOne{\textsuperscript{1}Physical Research Laboratory, Ahmedabad 380009, India.\\}
\affilTwo{\textsuperscript{2}IPAC, California Institute of Technology, MS 100-22, Pasadena, CA 91125, USA\\}
\affilThree{\textsuperscript{3}Division of Physics, Mathematics, and Astronomy, California Institute of Technology, Pasadena, CA 91125, USA\\}
\affilFour{\textsuperscript{4}STAR Institute, University of Li\`ege, All\'ee du 6 Ao\^ut 19c, 4000, Li\`ege, Belgium\\}
\affilFive{\textsuperscript{5}Physics Department, Auburn University, AL 368332, USA\\}


\twocolumn[{

\maketitle

\corres{aravind139@gmail.com}

\msinfo{}{}

\begin{abstract}
Jupiter Family Comets (JFCs), having orbital period less than 20 years, provide us with an opportunity to observe their activity and analyse the homogeneity in their coma composition over multiple apparitions. Comet 46P/Wirtanen with its exceptionally close approach to Earth during its 2018 apparition offered the possibility for a long-term spectroscopic observations. We used a 1.2 m telescope equipped with a low-resolution spectrograph to monitor the comet's activity and compute the relative abundances in the coma, as a function of heliocentric distance. We report the production rates of four molecules CN, C$_2$, C$_3$ and NH$_2$, and Af$\rho$ parameter, a proxy to the dust production, before and after perihelion. We found that 46P has a typical coma composition with almost constant abundance ratios with respect to CN across the epochs of observation. Comparing the coma composition of comet 46P during the current and previous apparitions, we conclude the comet has a highly homogeneous chemical composition in the nucleus with an enhancement in ammonia abundance compared to the average abundance in comets. 
\end{abstract}


}]


\doinum{12.3456/s78910-011-012-3}
\artcitid{\#\#\#\#}
\volnum{000}
\year{2023}
\pgrange{1--}
\setcounter{page}{1}
\lp{1}

\section{Introduction}\label{intro}
Comets are unpredictable Solar system bodies. The dynamical life of a comet is not a mere implication of its expected level of activity. Generally, Long-Period Comets (LPCs) show higher activity than Short Period Comets (SPCs). It is because the surface of the LPCs contains fresh material that has never been exposed or exposed very few times to Solar radiation compared to the processed surfaces of SPCs. In order to understand the overall activity of a comet during its apparition, extensive coverage along its orbit is necessary. This is important for SPCs to track the similarity/dissimilarity of their activities and relative abundance from previous apparitions. This could provide a great deal of information regarding the heterogeneity/homogeneity in the composition of the comet's nucleus over time. 

In this work, we discuss our long-term spectroscopic observations of 46P/Wirtanen (46P hereafter), belonging to the Jupiter Family \citep{KB2JFC_Levison97}. 46P was the original target of the \textit{Rosetta} mission after which it was changed to 67P/Churyumov–Gerasimenko due to certain delays in the launch \citep{46P_rosetta_target}. 46P, being a JFC has been observed during its previous apparitions \citep[eg., ][]{Ahearn_85, langland-shula, 46P_67P_2003Apparition, 46P_21year_waterprod, Lamy_46P_1998, 46P_farnham_schleicher_1998, Fink_1998, schulz_46P_1998}. However, the 2018 apparition presented the closest approach of the comet with Earth ($\sim$0.068 AU) in the last four centuries. This enabled us to probe the very interior of the coma even with seeing limited observations. The comet exhibited significant activity despite a 1.05 AU perihelion distance, and the hyperactivity of the comet was reported by \cite{46P_hypervolatiles} and \cite{46P_jehin_moulane}. The exceptional passage of 46P also resulted in the detailed study of the comet in imaging, IR/UV/optical spectroscopy, polarisation etc \citep[eg., ][]{46P_Farnham_jets_rotation, 46P_FUV, 46P_knight_schleicher, 46P_Neowise_CO_CO2, 46P_nowater_icegrains, 46P_photometry_polarimetry, 46P_rapidvariation_innercoma, 46P_sixoutbursts, 46P_polarisation, 46P_jehin_moulane}. Among the following sections, section \ref{obs_red} discusses the observation, data reduction, and analysis in detail, section \ref{discuss} discusses the various computed results, and section \ref{conclusion} summarises the work in brief.

\section{Observation and Analysis}\label{obs_red}
The availability of time on a telescope facility and compatible instruments does not guarantee the possibility of long-term coverage of a comet. The orbital parameters and Earth's orbital position play a major role in determining observability. The apparition of comet 46P in 2018 presented a prospect for long-term coverage in spectroscopy. As part of the long-term coverage, comet 46P was observed for 19 epochs spanning from 2018 October to 2019 February. The optical spectrograph LISA\footnote{\url{https://www.shelyak.com/produit/pf0021vis-lisa-slit-vis}} mounted on the 1.2 m telescope at Mt. Abu was used for all our observations of 46P. The details of the instrument are given by \cite{kumar_lovejoy}. The instrument provides an effective wavelength range of 3800–7000 Å with a resolving power of $\sim$800 and a pixel scale of $\sim 0.65$ arcsec/pixel. A long slit, oriented in the North–South direction, 3.8 arcmin in length and 3.6 arcsec in width was used for the observation of both the comet and standard star in all the epochs. During every epoch, the standard star HD 74721 was observed for flux calibration. Tungsten lamp spectra, zero exposure frames and ArNe lamp spectra were obtained for flat fielding, bias subtraction and wavelength calibration respectively. Also, the solar analog star, HD 19445 (G2V), was observed during every epoch in order to remove the continuum from the comet spectrum. 
\begin{figure}[h!]
	\centering
	\includegraphics[width=1\linewidth]{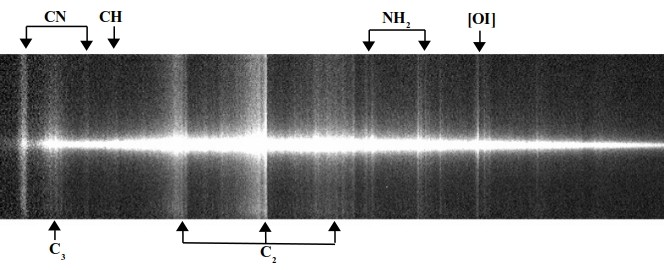}
    \caption{2D raw spectrum of comet 46P/Wirtanen observed on 2018-12-13 using the LISA spectrograph.}
    \label{fig:46P_raw}
\end{figure}
\begin{table*}[h!]
\centering
\setlength\tabcolsep{8pt}
\renewcommand{\arraystretch}{1.2}
\caption{Orbital parameters for the Comet 46P/Wirtanen at the time of observations}
\begin{tabular}{r r r r r c r}
\hline 
\hline
     & \multicolumn{1}{c}{{Heliocentric}}    & \multicolumn{1}{c}{Heliocentric} & \multicolumn{1}{c}{{Geocentric}} & \multicolumn{1}{c}{{Distance scale}} & \multicolumn{1}{c}{{Phase}}  & \multicolumn{1}{c}{{Exposure}}   \\
\multicolumn{1}{c}{{Date}}   &\multicolumn{1}{c}{{Distance} (r$_{H}$)} &
\multicolumn{1}{c}{velocity ($\dot{r_H}$)}&\multicolumn{1}{c}{{Distance} ($\Delta$)} & \multicolumn{1}{c}{{at photo-centre}} & \multicolumn{1}{c}{{angle}}  & \multicolumn{1}{c}{{Time}}     \\
\multicolumn{1}{c}{[UT]}    &  \multicolumn{1}{c}{[AU]}  & \multicolumn{1}{c}{[km s$^{-1}$]}& \multicolumn{1}{c}{[AU]} &    \multicolumn{1}{c}{[Km pixel$^{-1}$]} &  \multicolumn{1}{c}{[$^\circ$]}  & \multicolumn{1}{r}{{[seconds]}}   \\ \hline
2018-10-05 & 1.38 & -13.53 & 0.44 & 212 & 25.66 & 3600 \\
  2018-11-21 & 1.09 & -6.07 & 0.16 & 79 & 46.65 & 1800\\
  2018-11-28 & 1.07 & -4.19 & 0.13 & 62 & 46.03 & 1800\\
  2018-11-29 & 1.07 & -3.90 & 0.13 & 61 & 45.61 & 1800\\
  2018-11-30 & 1.07 & -3.62 & 0.12 & 58 & 45.09 & 900\\
  2018-12-08 & 1.06 & -1.30 & 0.09 & 43 & 36.07 & 1200\\
  2018-12-09 & 1.06 & -0.96 & 0.09 & 40 & 33.99 & 1200\\
  \hline
  2018-12-13 & 1.05 & 0.25 & 0.08 & 38 & 25.42 & 900\\
  2018-12-14 & 1.06 & 0.55 & 0.08 & 37 & 23.31 & 900\\
  2018-12-15 & 1.06 & 0.85 & 0.08 & 37 & 21.40 & 1200\\
  2018-12-27 & 1.07 & 4.36 & 0.10 & 49 & 27.01 & 900\\
  2018-12-28 & 1.08 & 4.64 & 0.11 & 51 & 28.13 & 900\\
  2019-01-11 & 1.13 & 8.15 & 0.18 & 86 & 33.41 & 1200\\
  2019-01-12 & 1.13 & 8.36 & 0.19 & 89 & 33.34 & 1800\\
  2019-01-13 & 1.14 & 8.58 & 0.19 & 92 & 33.24 & 1800\\
  2019-01-31 & 1.24 & 11.66 & 0.31 & 149 & 29.20 & 1200\\
  2019-02-01 & 1.25 & 11.81 & 0.32 & 153 & 28.93 & 1200\\
  2019-02-02 & 1.26 & 11.94 & 0.33 & 156 & 28.70 & 1200\\
  2019-02-03 & 1.26 & 12.06 & 0.33 & 160 & 28.48 & 1200\\
\hline
\multicolumn{7}{|l|}{The horizontal line separates the pre and post perihelion epochs  with perihelion passage on 2018-12-12.}\\
\hline
\end{tabular}
\label{observations_46P}
\end{table*}

\begin{figure*}[h!]
	\centering
	\includegraphics[width=0.9\linewidth]{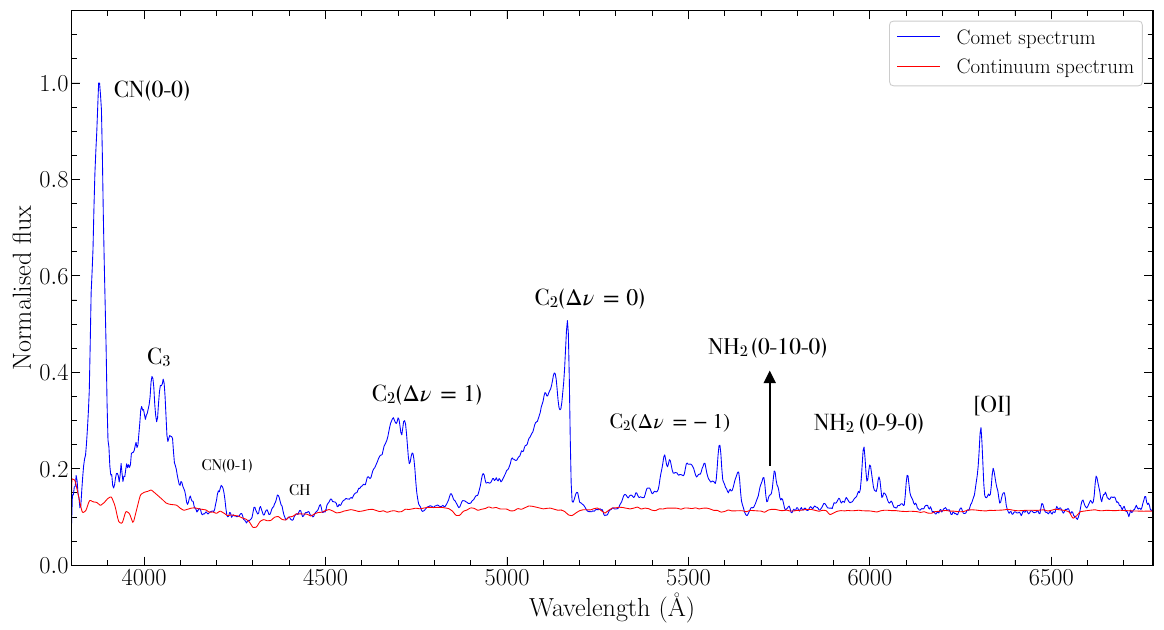}
    \caption{Optical spectrum of Comet 46P/Wirtanen observed on 2018-12-13 using the LISA spectrograph.}
    \label{fig:46P_continuum}
\end{figure*}

Figure \ref{fig:46P_raw} shows the raw 2D spectrum of the comet observed on 2018 December 13, with the detected emissions marked. The comet was observed using precise non-sidereal tracking available in the telescope control system. The NASA JPL HORIZONS\footnote{\url{https://ssd.jpl.nasa.gov/horizons.cgi}} service was used to generate the ephemerides for the comet during all the observation epochs. Since comets are extended objects, a separate sky frame of similar exposure to that of the comet, about 1 degree away from the photocentre, was obtained on all epochs. The brief details of the orbital parameters of the comet at the time of observation are given in Table.\ref{observations_46P}. 

The spectrum corresponding to the observations carried out on the 19 distinct epochs was extracted using Python scripts. The spectrum of the standard star observed on the same day is used to establish a reference trace for the instrument's spectrum. With necessary adjustments, this reference trace is applied to trace the comet's spectrum along the dispersion axis and thus extract its spectrum along the spatial axis.

The IRAF \textit{apall} task was employed for extracting the standard star spectrum as it allows for effective sky subtraction using regions on both sides of the target. Since similar narrow slit is used for both the comet and standard star observation, varying seeing conditions could result in the loss of flux from the standard star resulting in an overestimation in the flux of the comet while performing flux calibration. Hence, a slit correction factor, by making use of the Point Spread Function (PSF) along the slit length direction, as explained in \citet{slit_correction}, is introduced in order to cancel out this effect. Proper wavelength calibration and flux calibration were carried out using standard IRAF modules. Finally, prior to extracting the flux associated with each molecular emission and to mitigate the influence of the dust continuum, the underlying dust continuum is to be removed with the help of a solar analog spectrum. In this regard, initially, the calibrated solar analog spectrum is normalised and scaled to the comet continuum flux. A polynomial is fit to both the comet and solar analog spectrum for the continuum windows mentioned in \citet{C2019Y4_continuumremoval}. The scaled solar analog star spectrum is multiplied by the ratio of these polynomials (to correct for the redder nature of the cometary dust) in order to obtain a continuum spectrum of the comet. This continuum spectrum is now subtracted from the original comet spectrum to obtain the pure emission spectrum. The final calibrated spectrum of comet 46P for the observation on 2018 December 13, with the different emissions marked and dust continuum overplotted, is as shown in Figure \ref{fig:46P_continuum}. 


Strong emissions from CN$(\Delta \nu = 0)$, $\mathrm{C_3} (\mathrm{\lambda}4050$\AA) and $\mathrm{C_2} (\Delta \nu = 0)$ were observed on all epochs while emissions from NH$_2$ became evident as the comet approached perihelion.
Observations of the comet with the long slit in LISA were utilised to study the spatial variation in column density of various molecules (CN, C$_2$, C$_3$ and NH$_2$) in order to compute the evolution of their production rates as a function of heliocentric distance. Out of the different NH$_2$ bands, the NH$_2$ (0-10-0) (see Figure \ref{fig:46P_continuum}) was chosen due to its minimal blending with different C$_2$ bands. The production rate \textit{Q}, in molecules per second, for each molecule, is estimated using Haser model \citep{haser}, where minimum chi-square estimation between the observed column density and the theoretical column density computed using the equation mentioned in \cite{langland-shula} is employed. While the Haser model is usually fitted for nucleocentric distances beyond 10,000 km, due to the close approach of 46P, the Haser model has been fitted for nucleocentric distances varying between 2000-10000 km. The uncertainties in the production rate is estimated to be the systematic error computed from the minimum chi-square fitting. The further details regarding the method used to the observed column density profile to compute the production rates has been detailed in \cite{156P_aravind} and \cite{langland-shula}.

\begin{figure*}[h!]
	\centering
	\includegraphics[width=0.85\linewidth]{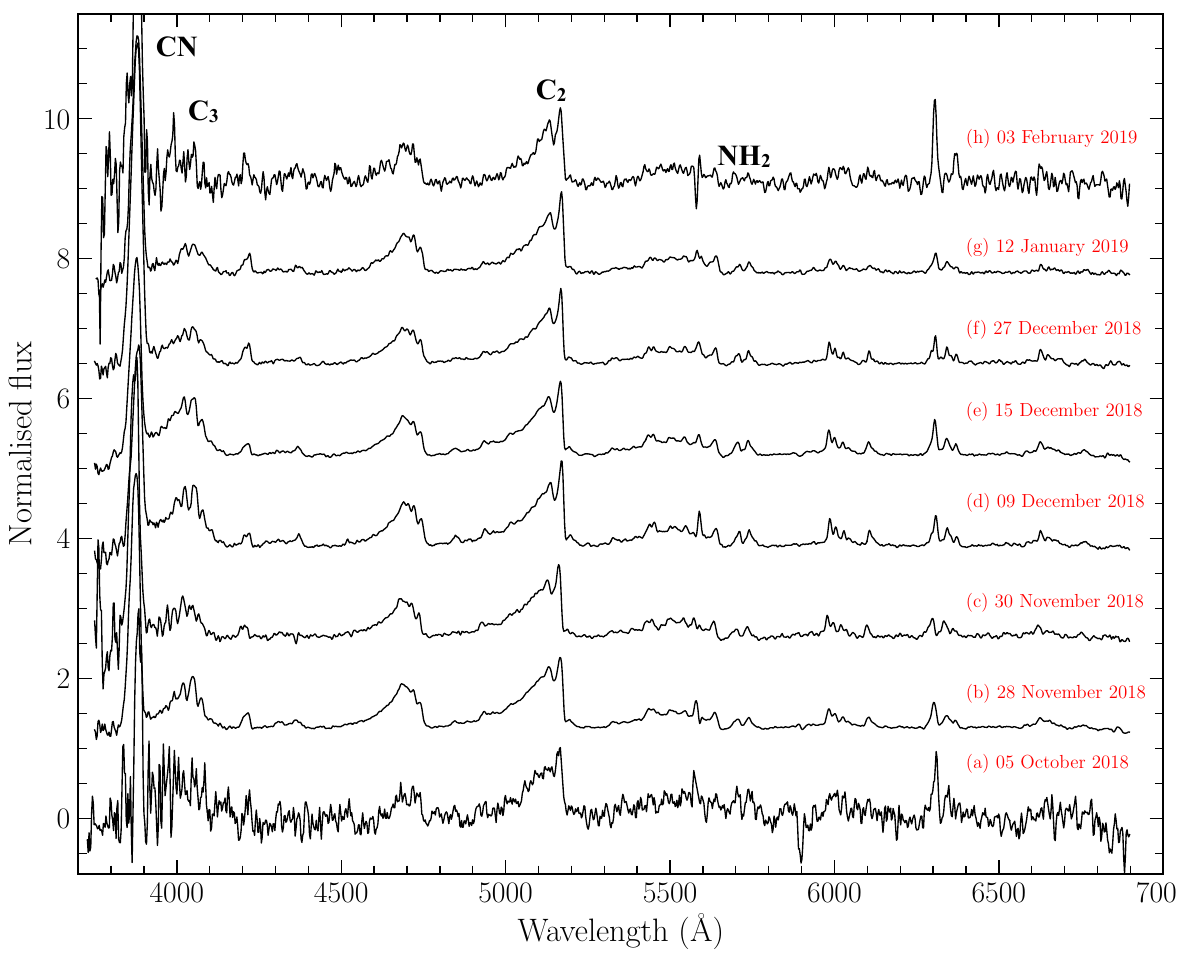}
    \caption{Optical spectra of comet 46P for selected observational epochs displayed with arbitrary offset in flux to illustrate the evolution of emissions.}
    \label{fig:46P_allspec}
\end{figure*}

The fluorescence efficiency (\textit{g}, ergs/molecule/second) at 1 AU for the molecules $\mathrm{C_2}$ and $\mathrm{C_3}$ have been taken from \citet{Ahearn_85}. Even though a set of revised fluorescence efficiency for the NH$_2$ emission bands have been reported in \cite{Kawakita_revised_g_NH2}, we have adopted the fluorescence efficiency for the (0-10-0) band to be half of the value reported in \cite{tegler_NH2}, similar to what has been used in \cite{Fink_1998} for the 1997 apparition of comet 46P and also in comet surveys reported by \cite{Fink_1996} and \cite{Fink_comet_survey}. 
Fluorescence efficiency for all molecules was scaled to $r_h^{-2}$ to determine the values at the corresponding heliocentric distance ($r_h$). Meanwhile, \cite{schleicher_CN_2010} have tabulated the g-factor of CN for different heliocentric distances and velocities. A double interpolation was performed on the provided table to obtain the exact \textit{g} values for the heliocentric distance and velocity at the time of observation. The scale lengths of the parent (l$_p$) and daughter (l$_d$) molecules for CN, C$_2$ and C$_3$ were taken from \cite{Ahearn_85}, while those for NH$_2$ were taken from \cite{cochran_30years} and were scaled to $r_h^{2}$.

\cite{Ahearn_Bowel_slope} defined a parameter Af$\rho$ as a proxy to the dust production present in the comet. Hence, the characteristic Af$\rho$ profile of the comet in Blue Continuum (BC) and Green Continuum (GC) narrow band filters \citep{HB} for various epochs can be used to study the variation in dust emission. In order to obtain the cometary flux within the band pass of the narrow band filters, the spectrum of the comet corresponding to each aperture is initially convolved with the transmission profile of the filter re-sampled to match the resolution of the instrument, which is later used to compute the magnitude and hence the Af$\rho$. The computed Af$\rho$ is further corrected with a phase function S($\theta$)\footnote{Composite Dust Phase Function for Comets:  \url{https://asteroid.lowell.edu/comet/dustphaseHM_table.txt}}, corresponding to the phase angle ($\theta$) at the time of observation, as defined in \citet{schleicher_phasefunction}. This method to compute the phase angle corrected A(0)f$\rho$ in different narrow band filter bandwidths from spectroscopic observation has been further detailed in \cite{156P_aravind}. The results obtained from these analyses for the comet 46P is discussed in the following section.

\section{Discussion}\label{discuss}

\begin{table*}[h!]
\centering
\setlength\tabcolsep{4pt}
\renewcommand{\arraystretch}{1.2}
\caption{{Activity of comet 46P at different epochs}}
\begin{tabular}{ccccccccc}
\hline
\hline
 \multicolumn{1}{c}{Date}    &\multicolumn{1}{c}{r{$_{H}$}}   & \multicolumn{1}{c}{$\Delta$} & \multicolumn{4}{c}{Log of the Production Rate (molecules per sec)}&\multicolumn{2}{c}{log[A(0)f$\rho$] (cm)}\\ 
 \multicolumn{1}{c}{(UT)} &   \multicolumn{1}{c}{(AU)} & \multicolumn{1}{c}{(AU)} & \multicolumn{1}{c}{CN} & \multicolumn{1}{c}{C{$_{2}$}({$\Delta \nu = 0$})} & \multicolumn{1}{c}{C{$_{3}$}} & \multicolumn{1}{c}{NH{$_{2}$}} &\multicolumn{1}{c}{BC} & \multicolumn{1}{c}{GC}\\ 
\hline
2018-10-05 & 1.38 & 0.44 & 24.54(0.06) & 24.62(0.08) & 23.56(0.24) &  -- & 1.75(0.32) & 1.72(0.36)\\
2018-11-21 & 1.09 & 0.16 & 25.06(0.05) & 25.13(0.06) & 24.16(0.19) & -- & 2.06(0.08) & 2.08(0.09)\\
2018-11-28 & 1.07 & 0.13 & 25.22(0.03) & 25.24(0.02) & 24.37(0.07) & 25.47(0.04) & 2.21(0.08) & 2.18(0.10)\\
2018-11-29 & 1.07 & 0.13 & 25.24(0.03) & 25.25(0.02) & 24.37(0.10) & 25.50(0.04) & 2.21(0.06) & 2.16(0.06)\\
2018-11-30 & 1.07 & 0.12 & 25.22(0.03) & 25.22(0.02) & 24.36(0.01) & 25.53(0.04) & 2.22(0.07) & 2.17(0.10)\\
2018-12-08 & 1.06 & 0.09 & 25.02(0.05) & 25.15(0.03) & 24.25(0.07) & 25.49(0.04) & 2.24(0.04) & 2.21(0.05)\\
2018-12-09 & 1.06 & 0.09 & 25.06(0.04) & 25.16(0.02) & 24.25(0.07) & 25.50(0.03) & 2.20(0.02) & 2.25(0.02)\\
\hline
2018-12-13 & 1.05 & 0.08 & 25.20(0.03) & 25.26(0.02) & 24.36(0.07) & 25.53(0.03) & 2.32(0.03) & 2.37(0.03)\\
2018-12-14 & 1.06 & 0.08 & 25.20(0.03) & 25.27(0.02) & 24.48(0.06) & 25.60(0.03) & 2.31(0.03) & 2.36(0.02)\\
2018-12-15 & 1.06 & 0.08 & 25.15(0.03) & 25.21(0.02) & 24.36(0.07) & 25.49(0.03) & 2.19(0.02) & 2.24(0.02)\\
2018-12-27 & 1.07 & 0.10 & 25.19(0.04) & 25.19(0.04) & 24.40(0.07) & 25.42(0.03) & 2.15(0.04) & 2.23(0.04)\\
2018-12-28 & 1.08 & 0.11 & 25.20(0.04) & 25.22(0.04) & 24.29(0.08) & 25.49(0.03) & 2.22(0.05) & 2.27(0.04)\\
2019-01-11 & 1.13 & 0.18 & 24.90(0.05) & 25.06(0.03) & 23.97(0.08) & 25.22(0.03) & 1.95(0.13) & 2.03(0.12)\\
2019-01-12 & 1.13 & 0.19 & 24.91(0.05) & 25.08(0.03) & 24.04(0.07) & 25.24(0.02) & 1.93(0.07) & 2.06(0.03)\\
2019-01-13 & 1.14 & 0.19 & 24.95(0.05) & 25.04(0.04) & 24.07(0.08) & 25.25(0.03) & 1.98(0.09) & 2.09(0.07)\\
2019-01-31 & 1.24 & 0.31 & 24.86(0.10) & 24.94(0.08) & 23.99(0.08) & 25.26(0.04) & 1.82(0.14) & 1.97(0.11)\\
2019-02-01 & 1.25 & 0.32 & 24.88(0.11) & 24.94(0.11) & 23.94(0.09) & 25.25(0.05) & 1.85(0.34) & 1.97(0.32)\\
2019-02-02 & 1.26 & 0.33 & 24.94(0.11) & 25.01(0.10) & 23.97(0.11) & 25.18(0.07) & 1.85(0.40) & 1.98(0.32)\\
2019-02-03 & 1.26 & 0.33 & 24.81(0.13) & 24.82(0.12) & 23.97(0.12) & 25.22(0.05) & 1.75(0.21) & 1.81(0.16)\\
\hline
\multicolumn{9}{|l|}{Values in the parenthesis represent the corresponding errors.}\\
\multicolumn{9}{|l|}{The horizontal line separates the pre and post perihelion epochs with perihelion passage on 2018-12-12.}\\
\hline
\end{tabular}
\label{result_table_46P}
\end{table*}

The latest apparition of 46P in 2018 presented a great opportunity to monitor various molecular emissions as the comet crossed perihelion. As the comet was approaching perihelion, the activity increased significantly resulting in the appearance of a large number of emission lines in the optical spectrum (see Figure \ref{fig:46P_allspec}). Detailed analysis of the evolution of individual emission lines can be performed to understand the comet composition in depth, which is beyond the scope of this work. The major emissions detected in the comet were the different bands of CN, C$_2$, C$_3$ and NH$_2$. 

As mentioned in \cite{Ahearn_85}, even though the production rate ratios or dust-to-gas ratio of a comet is not expected to vary significantly for minimal change in heliocentric distance, certain comets do exhibit such changes pointing to a possible heterogeneity in the composition of the comet's nucleus
\citep[eg., ][]{borisov_aravind}. Hence, extensive coverage of a comet along its orbit provides enough details to understand its basic compositional characteristics. Comparison of its activity and relative abundances of various molecules with previous apparitions also helps us confirm the homogeneity/heterogeneity of the nucleus composition \citep[eg., ][]{21P_schleicher}. The close geocentric approach of the comet during the current apparition implies that we are always looking at the innermost part of the coma where there can be drastic changes in activity due to the presence of jets or outbursts \citep{46P_Farnham_jets_rotation, 46P_sixoutbursts}. In such a scenario, imaging analysis would have an advantage over long-slit spectroscopy in investigating the outbursts and general production rate characteristics of the comet. The reason is imaging analysis would consider the emission arising from the total disk rather than a particular spatial axis, as in the case of long slit spectroscopy. Due to this disadvantage in spectroscopy, the presence of any localised jet or outburst varying from day to day at the location of the slit can affect the column density of the molecules and hence the computed production rates. This can be avoided to a larger extent by computing the column densities from the average flux obtained from both sides of the photocentre. However, while imaging is restricted to analysing only the emissions corresponding to the various filters available, spectroscopy is always advantageous in having a complete analysis of the different emissions occurring in the optical regime. Making use of this advantage, the NH$_2$ (0-10-0) emission band, which is not very well studied due to the absence of an NH$_2$ filter (NH narrow band filter is available as defined in \cite{HB}), has been analysed for the comet 46P. Analysing NH$_2$ abundance, the major daughter product of NH$_3$ \citep{shinnaka_2016_NH3_NH2, OPR_26comets, tegler_NH2}, is a shortcut to extract information on the ammonia abundance in the comet's nucleus.

\begin{figure*}[h!]
	\centering
	\includegraphics[width=0.8\linewidth]{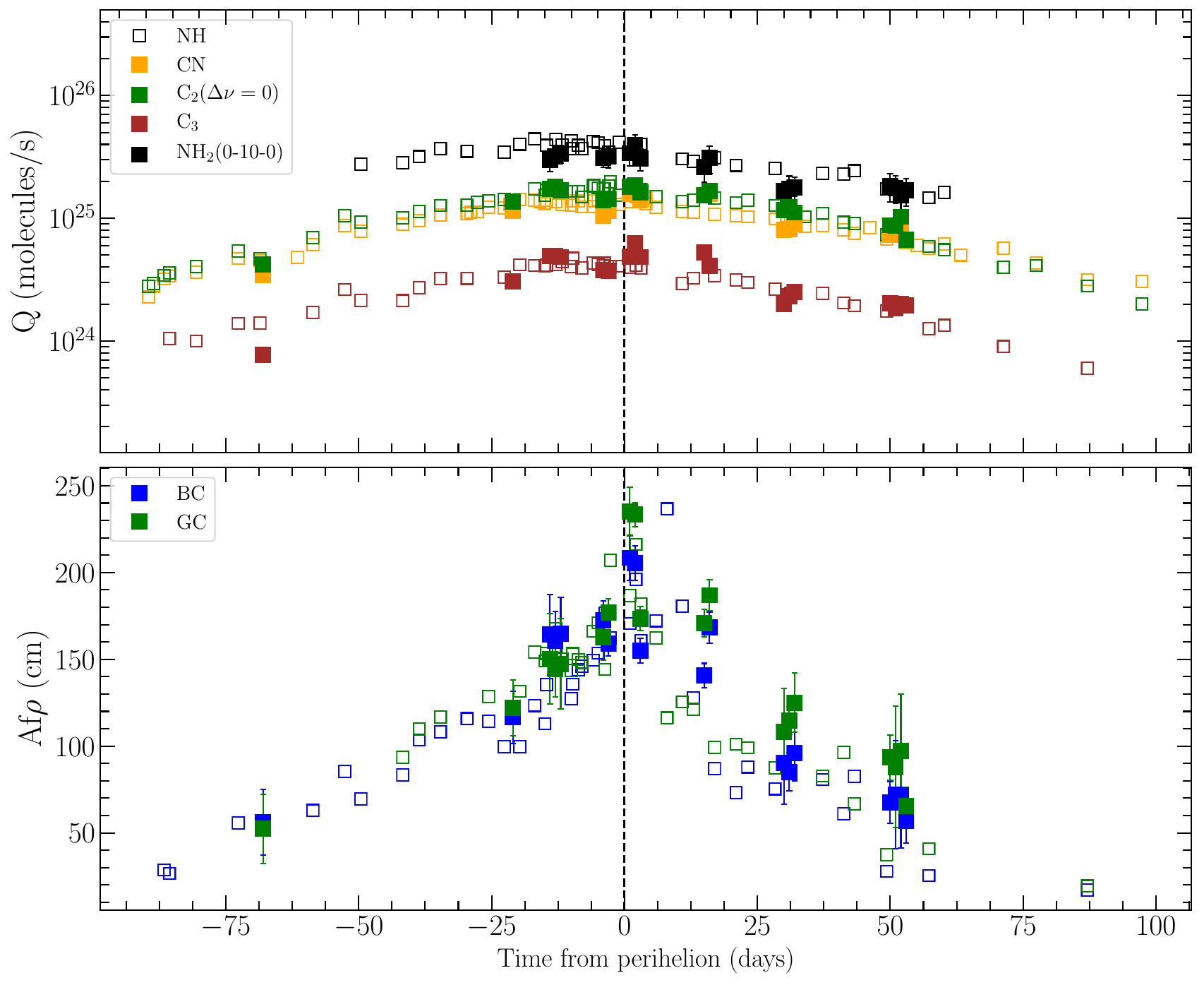}
    \caption{Observed trend in the computed production rates (CN, C$_2$, C$_3$, NH$_2$) and A(0)f$\rho$ (BC and GC). The filled squares represent the data from this work, and the open squares represent the data reported in \cite{46P_jehin_moulane}. The vertical dashed line marks the perihelion of the comet.} 
    \label{fig:46P_rates}
\end{figure*}

While \cite{46P_knight_schleicher} and \cite{46P_jehin_moulane} have reported the extensive coverage of comet 46P through photometric observations, we report the computed production rates of CN, C$_2$, C$_3$, NH$_2$, production rate ratios with respect to CN and the phase angle corrected A(0)f$\rho$ for two different continuum bands through long term spectroscopic observations. The upper panel of Figure \ref{fig:46P_rates} illustrates the variation in production rate of CN, C$_2$($\Delta \nu = 0$), C$_3$, and NH$_2$ (0-10-0) across perihelion (see Table \ref{result_table_46P} for corresponding values) along with a comparison with the production rates reported in \cite{46P_jehin_moulane}. It is seen that there is a gradual increase in the production rate of all four molecules as the comet approaches perihelion, after which it begins to drop. We note that the observed trend in the production rates of CN, C$_2$ and C$_3$ reported in this work are similar to those reported in \cite{46P_jehin_moulane} with slight variations at certain epochs. At the same time, \cite{46P_sixoutbursts} and \cite{46P_Farnham_jets_rotation} have reported multiple outbursts and a strong presence of CN jets as the comet approached perihelion. While \cite{46P_jehin_moulane} report the rotation period of the nucleus to be around 9 hours, they also mention that the CN jets do not rotate with time.
As previously mentioned, the comet's unprecedented close approach led us to investigate the activity in the innermost coma, where the photochemistry is likely highly complex, particularly due to the presence of strong jets and inner coma activity influenced by outbursts. Such combined effects could strongly affect the flux measured within the slit as the comet approached perihelion, causing the computed production rates to be slightly different than those obtained using photometry. 

\begin{figure*}[h!]
	\centering
	\includegraphics[width=1\linewidth]{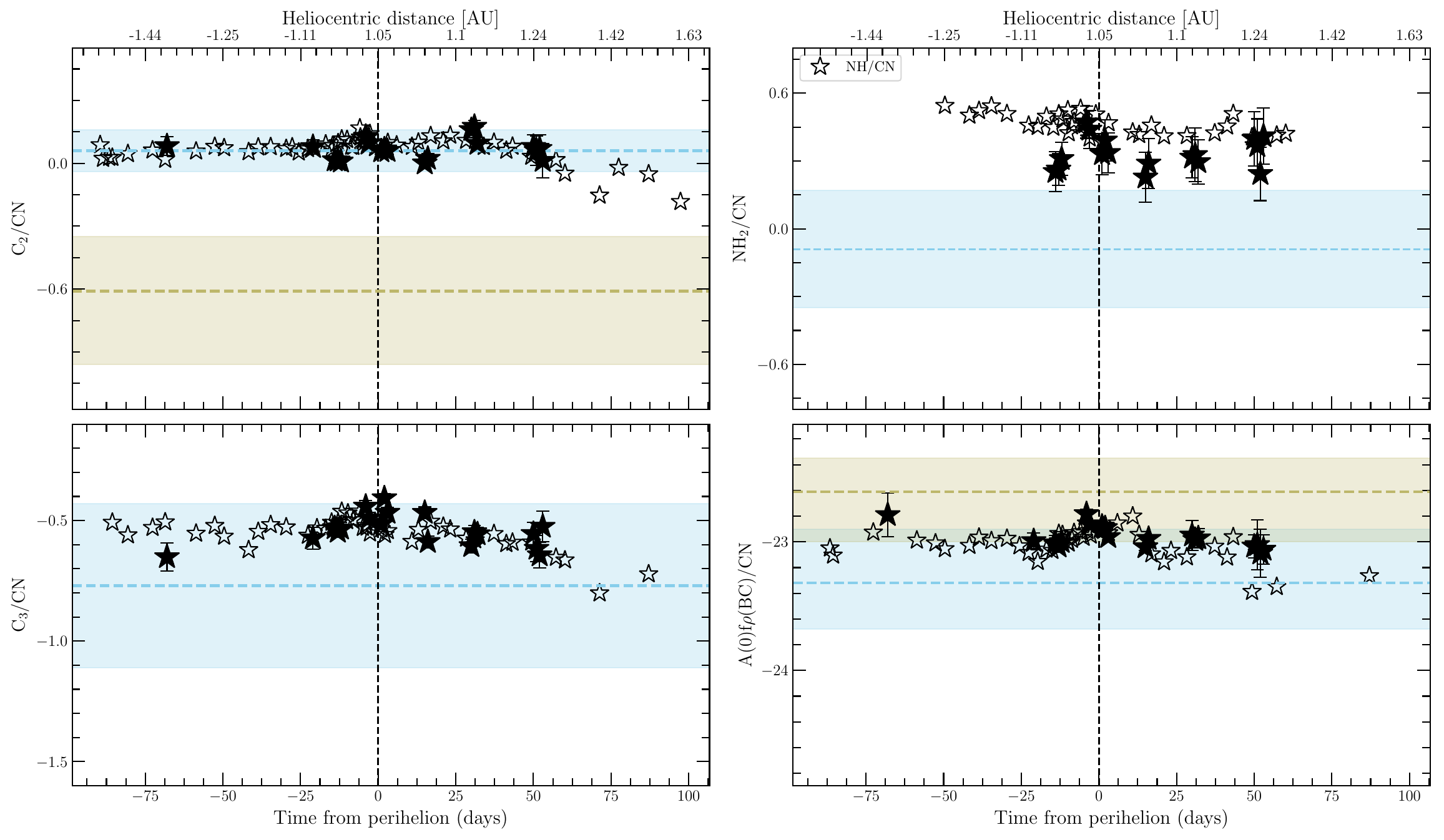}
    \caption{Observed trend in the computed production rate ratios and dust-to-gas ratio as a function of days to perihelion. The vertical dashed line represents the perihelion of the comet. The sky blue shaded region represents the mean value and range of ratios for comets with typical carbon composition, and the brown shaded region represents the same for comets with depleted carbon composition, as reported in \cite{Ahearn_85} and \cite{cochran_30years}. The filled stars represent the data from this work, and the open stars represent the data reported in \cite{46P_jehin_moulane}.}
    \label{fig:46P_ratios}
\end{figure*}

Even though there are reports of various outbursts and rapid variation of outgassing in the inner coma close to perihelion \citep{46P_rapidvariation_innercoma}, the activity for the molecules corresponding to these dates reported in this work and \cite{46P_jehin_moulane}, confirms that the outburst has not significantly affected the observed trends in the production rates or rate ratios. It can be seen from the left upper panel of Figure \ref{fig:46P_ratios} that the production rate ratio, Q(C$_2$)/Q(CN), is nearly constant across the observational epochs with a very good agreement to the trend reported in \cite{46P_jehin_moulane}, even though there was a slight difference in the production rates computed close to the perihelion. The computed abundance ratio, Q(C$_2$)/Q(CN), classifies the comet to be of typical carbon composition, as defined in \cite{Ahearn_85}, similar to what has been reported in \cite{46P_knight_schleicher}. In the same way, the abundance ratio Q(C$_3$)/Q(CN) (lower left panel in Figure \ref{fig:46P_ratios}) also lies in the region of comets with typical carbon composition, again with very good agreement to the ratio reported in \cite{46P_jehin_moulane}.

Interestingly, even though it has been widely stated that NH$_3$ dissociates into NH$_2$, which further dissociates into NH \citep[eg., ][]{cochran_30years, langland-shula, Brorsen_Teglar_NH3_NH2, wyckoff_tegler_NH2, 46P_jehin_moulane}, in the case of comet 46P, the computed production rates of both NH$_2$ and NH are seen to be almost similar across perihelion (see Figure \ref{fig:46P_rates}). Looking at the abundance ratios, comet 46P is seen to have an enhanced NH$_2$ composition with respect to CN (in comparison to comets reported in \cite{Fink_comet_survey} and \cite{cochran_30years}), which is also the same case with NH reported in \cite{46P_knight_schleicher} and \cite{46P_jehin_moulane} (see upper right panel of Figure \ref{fig:46P_ratios}). In addition, NH$_2$ being the direct daughter product of NH$_3$ with maximum yield \citep{OPR_26comets,shinnaka_2016_NH3_NH2, tegler_NH2} implies that the coma of 46P has an enhanced composition in ammonia. It is also worth noting that \cite{slanger_NH3_NH}, \cite{NH3_NH2_NH_TKSVega}, and \cite{Longslit_NH2_terrence} had put forward a possibility of NH$_3$ being a parent source of both NH$_2$ and NH. Such a scenario would create differences in the observed trends of Q(NH$_2$)/Q(CN) and Q(NH)/Q(CN).
In support of this, even though the production rate ratio of other molecules with respect to CN reported in this work and \cite{46P_jehin_moulane} are highly comparable, the ratio Q(NH)/Q(CN) displays a higher abundance in comparison to the ratio Q(NH$_2$)/Q(CN) pointing at a possibility of NH having parental species additional to NH$_2$. Keeping in mind that these trends cannot be confirmed with the observation of only one comet and that these trends could be influenced by the adopted scale lengths and fluorescence efficiencies for NH and NH$_2$, the possibility of both NH$_3$ and NH$_2$ being a parent source of NH cannot be ruled out.

The general dust activity in the comet along the orbit, inspected by analysing the variation in the A(0)f$\rho$ characteristic profile (radial profile of Af$\rho$ computed for increasing aperture size), corrected for phase angle, in the narrow band BC and GC filters, reveals a peak in the dust activity just one day after perihelion at a heliocentric distance of 1.05 AU, after which it drops significantly (see lower panel in Figure \ref{fig:46P_rates}), similar to the dust activity reported by \cite{46P_knight_schleicher}, \cite{46P_photometry_polarimetry}, and \cite{46P_jehin_moulane}. Again, it can be seen that despite the general trend in A(0)f$\rho$ computed for both BC and GC filter bands (as given in  Table \ref{result_table_46P}) being in good agreement to those reported in \cite{46P_jehin_moulane}, they are slightly on the higher side during certain epochs. As discussed before, this could be a direct effect of the activity in the slit being considered the activity in the entire coma. At the same time, the dust-to-gas ratio is not varying significantly for the observed range of heliocentric distance (see Figure \ref{fig:46P_ratios}) and is in very good agreement with the values reported in \cite{46P_jehin_moulane} (see lower right panel in Figure \ref{fig:46P_ratios}). In addition, the dust-to-gas ratio, (Af$\rho _{(BC)}$)/Q(CN), of comet 46P is observed to be normal (as defined by \cite{Ahearn_85}), pointing to a gas-rich coma.

\begin{figure*}[h!]
	\centering
	\includegraphics[width=0.9\linewidth]{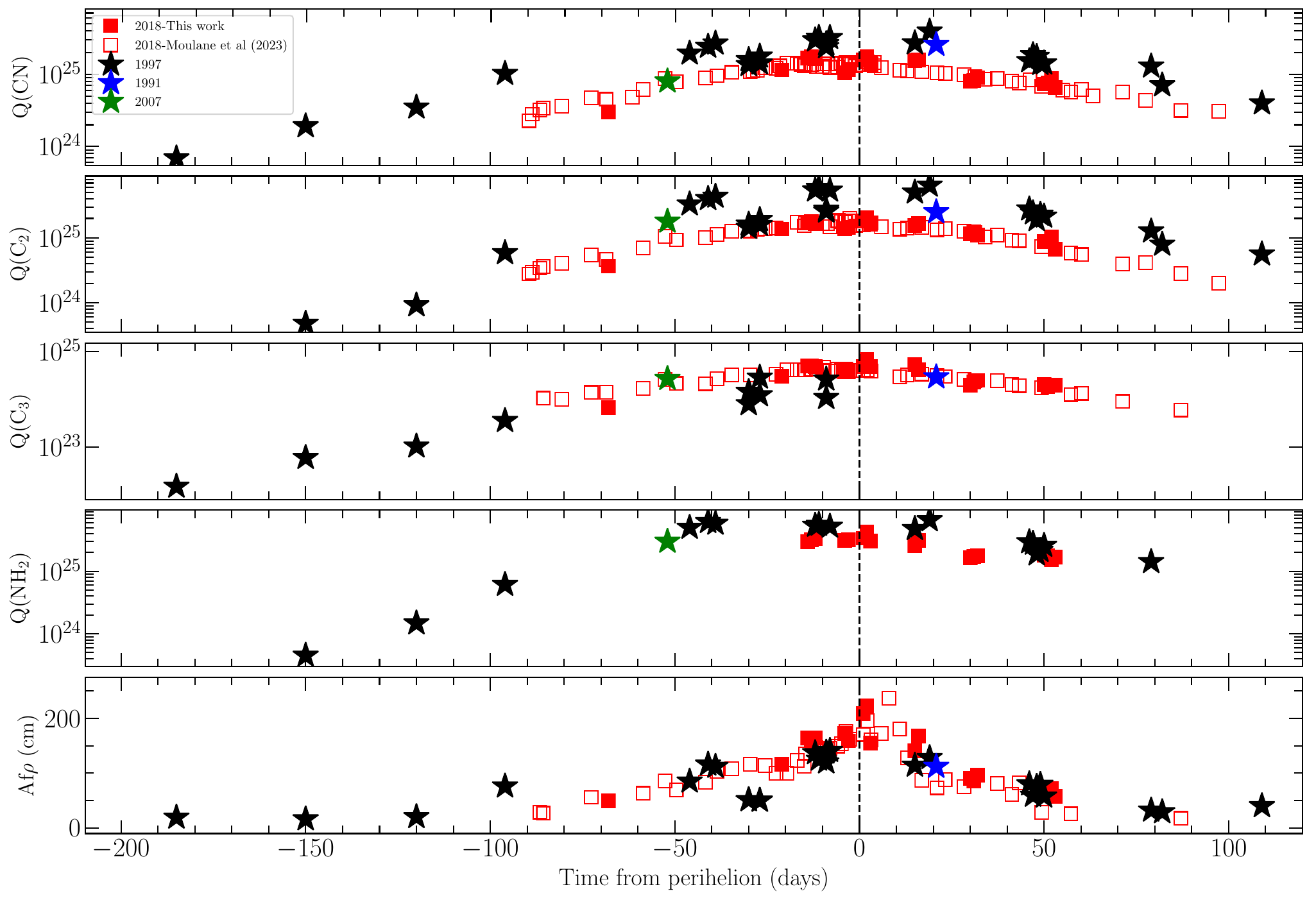}
    \caption{Comparison of the production rates of various molecules observed in comet 46P during the previous apparitions in 1991 \citep{46P_farnham_schleicher_1998}, 1997 \citep{46P_farnham_schleicher_1998, Fink_1998, schulz_46P_1998} and 2007 \citep{langland-shula}, with those reported in this work and \cite{46P_jehin_moulane} for the latest apparition in 2018.}
    \label{fig:46P_prev}
\end{figure*}

46P, being a short-period comet, has been observed via photometry and spectroscopy in its previous apparitions. In each of the previous apparitions, 1991, 1997, and 2007, where the comet was observed, 46P was reported to have a typical carbon-chain composition with a normal dust-to-gas ratio \citep{Ahearn_85, 46P_farnham_schleicher_1998, Fink_1998, schulz_46P_1998, Lamy_46P_1998, Jockers_46P_1998, 46P_67P_2003Apparition, langland-shula}. Figure \ref{fig:46P_prev} displays the comparison of gas and dust activity reported in this work to those reported for the previous apparitions of the comet. Comparing the 2018 apparition with the comet's 1997 apparition, covered extensively by \cite{46P_farnham_schleicher_1998}, \cite{Fink_1998} and \cite{schulz_46P_1998}, it is clearly seen that the outgassing has not changed significantly even after four apparitions. Reports by \cite{Fink_1998} during 46P's 1997 apparition also indicate a higher NH$_2$ abundance relative to CN, consistent with the findings in this study.

As mentioned in \cite{46P_jehin_moulane}, the observed abundance for different molecules for the latest apparition is also seen to be in good agreement with those reported for previous apparitions in 1991 \citep{46P_farnham_schleicher_1998, Ahearn_85}, 1997 \citep{46P_farnham_schleicher_1998, Fink_1998, schulz_46P_1998} and 2007 \citep{langland-shula}. Even with 46P having had a hyperactivity during its historic close flyby with Earth in this apparition, the similarity of coma composition with the previous apparitions implies a highly homogenous composition of the comet's nucleus. Additionally, the effects of domination by localised activity in the inner coma can be significant during such a close approach. Hence, further analysing the spectra of each individual epoch can reveal a great amount of detail regarding the variation in individual major and minor emissions from epoch to epoch during its hyperactivity.

\section{Summary and Conclusion}\label{conclusion}
In this work, we present the extensive spectroscopic observations of comet 46P/Wirtanen during its remarkable 2018 apparition, marked by an exceptionally close approach to Earth. The wide coverage in observational epochs helped us analyse the comet's activity as it went around the Sun. Using the observations from the low-resolution optical spectrograph, LISA, we derived multiple parameters that characterize the comet's activity and composition, including production rates, production rate ratios, A(0)f$\rho$, and the dust-to-gas ratio. Spectroscopy also facilitated the study of production rates of the NH$_2$ emission, a direct product of ammonia in the comet, which has lately not been very well studied due to the absence of dedicated filters. It has been observed that the production rates are more or less symmetric across perihelion while there is a clear asymmetry in the A(0)f$\rho$. The different abundance ratios classify the comet as a typical carbon-chain composition with a normal dust-to-gas ratio and slightly enhanced ammonia abundance.

The similarity of the reported coma composition of 46P in the current apparition to those in the previous apparitions, including the enhanced ammonia abundance, points to the comet's nucleus having a highly homogenous composition. Further dedicated studies focusing on the trends in the production of NH$_2$ and NH are required to understand better the parent sources responsible for these emissions.

In conclusion, the primary motive of this work, to analyse the general trend in the comet's activity as it crossed perihelion, helps us visualise the consistency of out-gassing in the comet across different apparitions. This study additionally furnishes evidence supporting the credibility of spectroscopic observations in analysing comets' activity trends. The results demonstrate a high level of agreement with photometric observations, even in the case of Comet 46P, which had an unprecedented close approach and exhibited hyperactivity.

In addition, taking into account that the minor species not discussed in this paper are also very significantly detected during many epochs of observations, the spectrum corresponding to each epoch can be further utilised along with fluorescence models to analyse the variation in these emissions (major and minor) from one epoch to another to understand better the photo-chemistry occurring in the coma of the comet 46P at different heliocentric distances. This collective knowledge can be effective in planning observations for the further detailed study of comet 46P during its future apparitions.

\appendix



\section*{Acknowledgements}

We acknowledge the local staff at the Mount Abu Observatory for their help that made these observations possible. Work at the Physical Research Laboratory is supported by the Department of Space, Govt. of India. 

Emmanuel Jehin is a FNRS Senior Research Associate. TRAPPIST is a project funded by the Belgian
Fonds (National) de la Recherche Scientifique (F.R.S.-FNRS) under grant T.0120.21.

This work is a result of the bilateral Belgo-Indian projects on Precision Astronomical Spectroscopy for Stellar and Solar system bodies, BIPASS, funded by the Belgian Federal Science Policy Office (BELSPO, Govt. of Belgium; BL/33/IN22\_BIPASS) and the International Division, Department of Science and Technology, 
(DST, Govt. of India; DST/INT/BELG/P-01/2021 (G)).
\vspace{-1em}


\bibliographystyle{aa}
\bibliography{reference}

\end{document}